\documentclass[12pt]{article}
\pdfoutput=1
\usepackage{color}
\usepackage{graphicx}
\usepackage{amsmath}
\usepackage{slashed}

\input xy
\xyoption{all}
\xyoption{web}

\newlength{\abstractwidth}

\renewcommand{\thefootnote}{\fnsymbol{footnote}}
\renewcommand{\thanks}[1]{\footnote{#1}}
\newcommand{\starttext}{
\setcounter{footnote}{0}
\renewcommand{\thefootnote}{\arabic{footnote}}}

\newcommand{\bea}{\begin{eqnarray}}
\newcommand{\eea}{\end{eqnarray}}
\newcommand{\ee}{\end{equation}}
\newcommand{\be}{\begin{equation}}

\newcommand{\ea}{\end{array}}
\newcommand{\bac}{\begin{array}{c}}
\newcommand{\bacc}{\begin{array}{cc}}
\newcommand{\barcl}{\begin{array}{r@{}c@{}l}}
\newcommand{\brcl}{\begin{array}{rcl}}
\newcommand{\bdm}{\begin{displaymath}}
\newcommand{\edm}{\end{displaymath}}

\def\det{{\rm det}}

\def\no{\nonumber}

\begin{document}
\starttext
\setcounter{footnote}{0}


\bigskip

\begin{center}

{\Large \bf A supersymmetric holographic dual of   a  fractional topological insulator}
\vskip 0.6in

{ \bf Martin Ammon and   Michael Gutperle}

\vskip .2in

{ \sl Department of Physics and Astronomy }\\
{\sl University of California, Los Angeles, CA 90095, USA}\\
{\tt \small ammon, gutperle@physics.ucla.edu; }

\end{center}

\vskip 0.2in

\begin{abstract}

We construct a supersymmetric generalization of the holographic dual of  a fractional topological insulator found in \cite{HoyosBadajoz:2010ac}. This is accomplished   by introducing a nontrivial gauge field on the world volume of the probe D7 brane. The BPS equations are derived from the $\kappa$-symmetry transformation of the probe brane. The BPS equations  are shown to reduce to two first oder nonlinear partial differential equations. Solutions of the BPS equations correspond to a probe brane configuration which preserves four of the thirty-two supersymmetries of the $AdS_5\times S^5$ background.
Solutions of the BPS equations which correspond to a holographic  fractional topological insulator  are obtained numerically.

\end{abstract}

\newpage


\baselineskip=16pt
\setcounter{equation}{0}
\setcounter{footnote}{0}

\newpage

\section{Introduction}
\setcounter{equation}{0}
\label{sec1}

The AdS/CFT correspondence, and more generally gauge/gravity dualities, provide a powerful 
tool for studying strongly coupled field theories in states with finite density, and hence might be useful in condensed matter physics (see \cite{Hartnoll:2009sz,Herzog:2009xv,McGreevy:2009xe} and references therein). For example, superfluids \cite{Hartnoll:2008vx,Gubser:2008wv,Horowitz:2010gk} and Non-Fermi liquids \cite{Liu:2009dm} can be realized in bottom-up models within this framework. Note that such  bottom-up models are generically formulated in terms of four or five dimensional gravity theories coupled to scalars and gauge fields.  It might or might not be possible to embed such  models  into  string theory. Consequently, the dual field theory  formulation in terms of elementary fields and a Lagrangian is in general not known or  might even not exist. String theory embeddings of these bottom-up models can be realized by using probe branes in the background of D3--branes. In this case the dual field theory is known explicitly. For example p-wave superfluids \cite{Ammon:2008fc,Basu:2008bh,Ammon:2009fe}, Fermi surfaces \cite{Ammon:2010pg} and the gravity dual of a Quantum Hall Plateau transition \cite{Davis:2008nv} were investigated.

In this paper we consider the dual gravity description of a fractional topological insulator. A (conventional) topological insulator is  a proposed new type of quantum matter  which is not adiabatically connected to an ordinary insulator.  They are characterized by fully gapped excitations in the bulk and gapless boundary modes, whose vanishing mass is protected by a discrete symmetry.  Topological insulators have been a very active field of research in the past couple of years (see \cite{kana,Hasan:2010xy,moore1,qi1} for reviews with  references to the original literature). 

A very simple model for a $Z_2$ time reversal invariant topological insulator \cite{Qi:2008ew}  is given  by a Dirac fermion in 3+1 dimensions with a spatially varying mass
\be\label{massdirac}
{\cal L} = \bar \psi \big(i \slashed{\partial}- m (x) \big) \psi.
\ee
One considers an interface across which $m(x)$  jumps from a positive to a negative real value.  For such an interface there is are  massless  localized  fermionic degrees of freedom \cite{Jackiw:1975fn,Callan:1984sa}, realizing the gapless boundary mode of the topological insulator. The interface separates topological trivial and nontrivial phases of a $Z_2$ topological invariant of the electron system (such as the $\theta$ term $\theta \, \bf{E} \cdot \bf{B}$ where $\theta=0$ corresponds to the trivial $Z_2$ insulator while $\theta = \pi$ corresponds to the $Z_2$ non--trivial insulator). 

By analogy to the relation of the integer and fractional quantum hall effect we can ask whether it is possible to construct a time reversal invariant fractional topological insulator for which $\theta$ is a non--integer fractional multiple of $\pi.$ In three spatial dimensions such a realization of a fractional topological insulator is known \cite{Maciejko:2010tx} (see also \cite{Swingle:2010rf,Maciejko:2011ed}). The idea is that the charge carriers, i.e. electrons, are made up of partons which carry fractional charge. To ensure that the partons confine into electrons outside the topological insulator we have to add a statistical gauge field which is  deconfined inside a topological insulator. 

The structure of this paper is as follows. In section \ref{sec2a} we review the probe brane setup for  a fractional topological insulator following \cite{HoyosBadajoz:2010ac}. Such a system breaks all supersymmetries due to the position dependent mass. In section \ref{sec2} we present a field theory analysis that identifies a counterterm which restores some of the broken supersymmetries  of the topological insulator. It is shown that the introduction of such a counterterm  can be 
engineered by turning on a specific gauge field on the D7 brane probe, emulating ideas which were used in the construction of supersymmetric interface theories \cite{D'Hoker:2007xy,D'Hoker:2007fq,D'Hoker:2008wc,Chiodaroli:2009yw}.
 In section \ref{sec3} we construct the BPS equations from the $\kappa$-symmetry transformations of the probe D7 brane. The conditions  that four of the thirty-two supersymmetries of the $AdS_5\times S^5$ vacuum are preserved  
 reduce to two first order nonlinear partial differential equations\footnote{Note that a supersymmetric probe D7 brane preserves at most sixteen of the thirty two supersymmetries of the vacuum.}. In section \ref{sec4} we obtain solutions using 
 various techniques. The solution which has the right asymptotic behavior to describe a topological insulator and is everywhere regular is obtained 
 numerically. We discuss possible applications and generalizations of our 
 results in section \ref{sec5}. Conventions, review material and technical details of the calculations are relegated to various appendices.

\section{A holographic fractional topological insulator} \label{sec2a}
In \cite{Maciejko:2010tx} (see also \cite{Swingle:2010rf,Maciejko:2011ed})  the idea of a fractional topological insulator was introduced. In this model the charge carriers fractionalize into partons with fractional charge.  Consequently, while  the system is still time reversal invariant, the topological invariant does not have to be $Z_2$ valued any more. Additional degrees of freedom (i.e. a strongly interacting gauge field coupled to the partons) ensure that outside the topological insulator the partons are confined and only integer charged electrons appear.

A holographic realization of a fractional topological insulator was presented in  \cite{HoyosBadajoz:2010ac,Karch:2010mn}, where the partons are realized by adding matter to a $\mathcal{N}=4$ SYM "phonon" bath.  We give a brief review of this construction following \cite{Karch:2003nh}. The starting point is the intersection of D3 and D7 branes in ten dimensional flat spacetime (with coordinates $X_0, X_1, \dots, X_9$).

\begin{table}[htdp]
\begin{center}
\begin{tabular}{|c|c|c|c|c|c|c|c|c|c|c|}
\hline
&0&1&2&3&4&5&6&7&8&9\\
\hline
D3&$\bullet$ &$\bullet$&$\bullet$&$\bullet$&&&&&&\\
\hline
D7&$\bullet$&$\bullet$&$\bullet$&$\bullet$&$\bullet$&$\bullet$&$\bullet$&$\bullet$&&\\
\hline
\end{tabular}
\end{center}
\label{default}
\caption{D3/D7 system. The bullets denote Neumann boundary conditions. The D7 brane can be separated from the D3 in the 8-9 directions.}
\end{table}

Whe choose the $N_c$ coincident D3 branes to be located at $X_4 = \dots = X_7 = X_8=X_9=0.$ The massless modes in the D3-D3 string sector give rise to an $\mathcal{N}=4$ supersymmetric Yang-Mills theory with gauge group $SU(N_c).$ From the point of the fractional topological insulator, these supersymmetric gauge theory degrees of freedom should be viewed as  "phonons". 

Moreover, we also add fundamental matter degrees of freedom to our setup by embedding $N_f$ coincident D7 branes. The D7 branes are aligned along the $X_0, X_1, \dots, X_7$ direction. The world volume coordinates $x_a$ ($a \in \{ 0,1,\dots, 7 \}$) can therefore be identified with $X_a,$ i.e. $x_a = X_a$. If the D7 branes are located at $X_8=X_9=0,$ the lowest energy excitations of the D3-D7 strings describe massless matter. The symmetries of this configuration are $SO(4)=SU(2)_R\times SU(2)_L$ rotations in the $X_4$-$X_7$ space and $SO(2)=U(1)_R$ rotations in the $X_8-X_9$ plane. The massless modes in the D3-D7 string sector can be written in terms $\mathcal{N}=2$ hypermultiplets where the $SU(2)_R\times U(1)_R$ rotations correspond to the superconformal R-symmetry. Moving the location of D7 brane away from $X_8=X_9=0$  breaks the $U(1)_R$ symmetry and introduces mass terms for the hypermultiplets. 

In the large $N_c$ limit, for $\lambda = N_c g_{YM}^2 \gg 1,$ the $N_c$ coincident D3 branes can be replaced by the $AdS_5\times S^5$ geometry, which is conveniently expressed as follows
\be\label{adsmeta}
ds^2= {r^2+ \rho^2\over R^2} (-dt^2+\sum_{i=1}^3 dx_i^2) +  {R^2 \over r^2+\rho^2} \Big( dx_4^2+dx_5^2+ dx_6^2+ dx_7^2+ dX_8^2 + dX_9^2),
\ee
where
\be
r^2= x_4^2+x_5^2+x_6^2+x_7^2, \quad \quad \rho^2=X_8^2+X_9^2.
\ee 
In these coordinates the $AdS_5$ boundary is located at $r\to \infty$. $R$ is the curvature radius of $AdS_5$ and is given by $R^4 = \lambda \, \alpha^{\prime \, 2}.$ From now on we will set $R=1$ and therefore $\alpha^{\prime \, -2} = \lambda.$

Let us now embed the $N_f$ coincident D7 branes. In the present paper we will only consider the the probe limit (i.e $N_f \ll N_c$), where the back-reaction of the D7 brane can be neglected. Then  the flavor degrees of freedom are described by embedding the D7 brane world volume  along an  (asymptotically) $AdS_5\times S^3$  subspace inside the $AdS_5\times S^5,$ with world volume coordinates $x_a.$ 

Without loss of generality, let us single out the field theory direction $x_3$ and consider a non-trivial profile $m(x_3)$ for the mass of the hypermultiplets. On the gravity side, this corresponds to a non-trivial embedding of the D7 branes in the transverse $(X_8, X_9)$ space of the form
\be\label{ansatzX8}
X_8= X_8(r,x_3), \quad X_9=0.
\ee
Since $X_8 \neq 0,$ the $U(1)_R$ symmetry is broken. In principle, $X_8$ can be any function of $x_a$ with $a\in \{0, \dots, 7\}.$ In order to preserve the $SO(4) = SU(2)_L \times SU(2)_R$ symmetry $X_8$ depends only on $r$ and not on $x_a$ with $a\in \{ 4,5,6,7 \}$ explicitly. The mass $m(x_3)$ can be read off from the boundary behavior of $X_8(r,x_3)$
\be\label{limit}
\lim_{r\to \infty} X_8(r, x_3) = m(x_3) + \mathcal{O}\left({1\over r^2}\right).
\ee    
Besides the scalar functions $X_{8,9}$ there are also gauge fields living on the D7 branes. In case of $N_f$ coincident D-branes the gauge fields are valued in $U(N_f).$ The dynamics of $N_f$ coincident D7 branes is given in terms of a Dirac-Born-Infeld (DBI) term, as well as a Wess-Zumino (WZ) term. As we will be interested only in the U(1) part of the $U(N_f)$ world-volume gauge fields and scalars, the relevant part of their action reads
\be
S_{D7} = S_{DBI} + S_{WZ} 
\ee
with
\be
S_{DBI} = - N_f \mu_7 \int d^8 x \sqrt{- \, \mbox{det}\, (G_{ab}+ F_{ab}) } ,
\ee
where $F_{ab}$ is the Abelian field strength tensor\footnote{In order to restore the correct $2\pi\alpha^\prime$ factors in the DBI and WZ action we have to rescale $F \rightarrow 2\pi\alpha^\prime F$} and $G_{ab}$ is the induced metric defined by\footnote{We do not distinguish between $x_a$ and $x^a$, i.e. $x_a = x^a$.}
\be
G_{ab} = {\partial X^\mu\over  \partial x^a}  {\partial X^\nu\over \partial x^b}  g_{\mu\nu}.
\ee
The Wess-Zumino part of the D7 brane action, $S_{WZ},$ is given by
\be\label{WZac}
S_{WZ} = N_f \mu_7 \int \mathcal{P}[ \sum_p C_{(p)}] \wedge e^{F} ,
\ee
where $\mathcal{P}[C_{(p)}]$ denotes the pull back of the background p-form field $C_{(p)}$ and $e^F = 1 + F + 1/2 F \wedge F + \dots$. The integral in (\ref{WZac}) singles out the correct p-form, i.e. in our case a 8-form with legs along the world volume coordinates.
  
For the embedding considered in (\ref{ansatzX8}) with all field strength tensors $F_{ab}=0,$ the WZ action does not contribute and the DBI action $S_{DBI} = - N_f \mu_7 \, \mbox{vol}(S^3) \int\! d^4x \,dr \, {\cal L}$ reads
\be\label{lagtop}
 {\cal L}= \sqrt{-det(G_{ab}) } = r^3 \sqrt{ 1+  (\partial_r X_8)^2 + {1\over (r^2 +X_8^2)^2} (\partial_{x_3} X_8)^2}.
\ee

A holographic topological insulator is given by a solution of the equations of motion following from the Lagrangian (\ref{lagtop}) with the following boundary condition  in the asymptotic AdS region
\be\label{limitx8}
\lim_{r\to \infty} X_8(r, x_3) = M_0   {x_3\over |x_3|} + \mathcal{O}\left({1\over r^2}\right).
\ee

Since $X_8$ is interpreted as the mass of the flavor degrees of freedom  is can be identified with the position dependent mass $m(x)$ for the fermions   (and their superpartners) in (\ref{massdirac}).  The equations of motion are quite complicated and an exact solution with these boundary conditions is not known at present since.  In  \cite{HoyosBadajoz:2010ac} numerical solutions of the equations of motion were found using the heat method.

\section{Ansatz for a supersymmetric topological insulator}
\setcounter{equation}{0}
\label{sec2}

In this section we  show  that  the  position dependent mass of  the field theoretic topological insulator model presented in the previous section breaks all supersymmetries. However adding a localized counterterm can restore some of the supersymmetries. This argument  motivates the  modification of the holographic model  by introducing a nontrivial gauge field on the probe D7 brane.

\subsection{Field theory analysis}\label{ftanalys}

The low energy theory of the D3/D7 brane intersection considered in the previous section is given by $\mathcal{N}=4$ supersymmetric Yang-Mills theory with gauge group $SU(N_c)$ coupled to $N_f$ $\mathcal{N}=2$ supersymmetric hypermultiplets. The $\mathcal{N}=4$ supersymmetric vector multiplet -- describing the low energy theory of strings ending on the $N_c$ D3-branes -- can be decomposed into one $\mathcal{N}=1$ vector multiplet $V_\mu$ and three $\mathcal{N}=1$ chiral multiplets called $\Phi_1, \Phi_2$ and $\Phi_3.$ The lowest components of the chiral multiplets, i.e. the scalar fields $\phi_1, \phi_2$ and $\phi_3$ can be identified with the six coordinates $X_4, X_5, \dots, X_9$ transverse to the D3-branes: $\phi_1 \sim X_4 + i X_5, \phi_2 \sim X_6 + i X_7$ and $\phi_3 \sim X_8 + i X_9.$ 

The low energy theory of the strings connecting the $N_c$ D3-branes and the $N_f$ D7 branes are given in terms of $N_f$ $\mathcal{N}=2$ supersymmetric hypermultiplets. For simplicity, we restrict ourselves to the case $N_f=1$ in the following discussion. An $\mathcal{N}=2$ supersymmetric hypermultiplet contains four fermionic and four bosonic degrees of freedom. It can be decomposed into an $\mathcal{N}=1$ chiral multiplet $Q$ and an $\mathcal{N}=1$  anti-chiral multiplet $\tilde Q^\dagger$, or equivalently into two $\mathcal{N}=1$ chiral multiplets $Q$ and $\tilde{Q}.$ The scalar fields of these chiral multiplets are denoted by $q$ and $\tilde{q},$ respectively, while the fermions are denoted by $\psi$ and $\tilde{\psi}.$  

In the  $\mathcal{N}=1$ language, the interactions of the vector and hypermultiplets are specified by  the superpotential $W$
\be\label{superpot}
W = \epsilon^{ijk} \mbox{tr} \left( \Phi_i \Phi_j \Phi_k \right) + \tilde{Q} (m + \Phi_3) Q.
\ee
In the following only  the superpotential term  involving the mass of the quarks,  $m,$  will be important. Although not obvious in the $\mathcal{N}=1$ superspace language, the field theory is $\mathcal{N}=2$ supersymmetric and has a $SU(2)_\mathcal{R}$ $R$ symmetry as well as a global $SU(2)_L$ symmetry. Note that both symmetries are also present in the brane intersection giving rise to the $SO(4) = SU(2)_L \times SU(2)_\mathcal{R}$ rotational invariance in the $(X_4, X_5, X_6, X_7)$ space. Note that the pair $(q, \tilde{q}^\star)$ transforms as $(0, 1/2)$ under $SU(2)_L \times SU(2)_\mathcal{R}$ while the fermions $(\psi, \tilde{\psi}^\dagger)$ are singlets under $SU(2)_L \times SU(2)_\mathcal{R}.$

The realization of the topological insulator makes the mass of the fermions dependent on one of the spatial coordinates, where we choose $x_3$ without 
loss of generality. This can be achieved by replacing $m$ by $m(x_3)$ in the superpotential (\ref{superpot}), introducing a  position dependent masses for 
both the fermions and the scalar superpartners. It was shown  \cite{Clark:2004sb}  that position dependent couplings  in the superpotential break all of the  supersymmetries since the supersymmetric variation of the Lagrangian is no longer a total derivative. 

In \cite{Clark:2004sb} is was also shown that half of the original supersymmmetries can be restored by adding a counter-term  to the action of the form (we give a review of the argument in appendix \ref{appendd})
\be\label{counter2}
{\cal L}\to {\cal L} + \Delta {\cal L},
\ee
where $\Delta {\cal L}$ is given by
\bea\label{counter1}
\Delta {\cal L} &=& - 2 \ \mbox{Im} \, \left({\partial m \over \partial x_3} \ {\delta W\over \delta  m(x_3)} \right) \\ \nonumber
&=& i {\partial m\over \partial x_3} \left\{ \left(  {\delta W\over \delta  m(x_3)} \right)- \left({\delta W\over\delta m(x_3)}\right)^*\right\}.
\eea
In the last step of equation (\ref{counter1}) we assumed that $m(x_3)$ is real. Note that if the position dependence is of the form (\ref{limitx8}), i.e. $m(x_3)=M_0 \,x_3/|x_3|$, the counterterm (\ref{counter1}) will be delta-function localized at the interface location $x_3=0$.
For a real mass $m(x_3)$ and for the superpotential (\ref{superpot}) the counterterm (\ref{counter1}) becomes
\be\label{counterb}
\Delta {\cal L} = i {\partial m\over \partial x_3}   \Big (  Q \tilde{Q}  - Q^\star \tilde{Q}^\star \Big).
\ee
Let us compose the scalar fields $q$ and $\tilde{q}$ into a vector $\hat{q}$ (whose components are denoted by $\hat{q}^m$ with $m=1,2$)
\be
\hat{q} = \left( \begin{array}{c} q \\ \tilde{q}^\star \end{array} \right).
\ee
We can write the part of the counterterm quadratic in the scalar fields $q$ and $\tilde{q}$ as
\bea\nonumber
\Delta{\cal L} &=& i {\partial m\over \partial x_3} \left( \tilde{q} q - q^\star \tilde{q}^\star \right)\\ \label{counter3}
&=& i {\partial m\over \partial x_3} \hat{q}^\dagger \sigma^2 q + \dots ,
\eea
where $\hat{q}^\dagger = \hat{q}^{\star \, T} = (q^\star, \tilde{q})$ and $\sigma^I$ are the Pauli matrices. The terms represented by $\dots$ in equation (\ref{counter3}) are quadratic in the fermions $\psi$ and $\tilde{\psi}.$ 

Recall that field theory has $SU(2)_L\times SU(2)_{\mathcal{R}}$ symmetry for a massive hypermultiplet and $\hat{q}$ transforms under $(0,1/2).$ Hence the counterterm (\ref{counter3}) has the following important properties: It  has scaling dimension $\Delta=2$ and transforms as $(0,1)$ under the $SU(2)_L\times SU(2)_{\mathcal{R}}$ symmetry.

The two components of the four  $\mathcal{N}=1$ supersymmetry which is preserved by the addition of the counterterm is given by (\ref{susyproj}).  Since our field theory model is really $\mathcal{N}=2$ supersymmetric the counterterm preserves four supersymmetries.

\subsection{Modified holographic ansatz }
The field theory analysis showed that in order to restore some supersymmetry a counterterm has  to be introduced. In this section we present  an ansatz for the holographic realization of such a counterterm. We need to identify a world 
volume field which can reproduce (\ref{counter1}) at the AdS boundary. The counterterm  corresponds to an operator of dimension $\Delta=2$ and it transforms under the    $SU(2)_L\times 
SU(2)_{\mathcal{R}}$ symmetry  in the  spin $(0,1)$ representation.  In \cite{Kruczenski:2003be} the excitation spectrum on the probe D7 brane was analyzed in detail. It was 
shown an operator with these properties is dual to a KK excitation of the gauge 
field called $\phi_I^-$. In the following we give an explicit realization of that particular gauge field excitation.

As we shall show the counterterm can be realized by turning on a world volume gauge field $A$ which gives rise to a non-vanishing field strength tensor with components $F_{ab} = \partial_a A_b - \partial_b A_a$. The ansatz for the gauge field we choose is given by
\bea\label{gaugeansatz}
A=
\left(
\begin{array}{c}
  A_4\\
  A_5\\
  A_6\\
  A_7
\end{array}
\right) ={f(x_3 ,r)\over r^2}\left(
\begin{array}{c}
  x_5\\
 -  x_4\\
  -x_7\\
   x_6
 \end{array}
\right)= {f(x_3 ,r)\over r^2}\left(
\begin{array}{c}
  y_4\\
  y_5\\
  y_6\\
   y_7
 \end{array}
\right).\label{gaugef}
 \eea
In the second equality we have defined a second set of coordinates $y_i$ will be useful in the calculation of the BPS projectors later on. Note that the dependence of $A$ on the coordinates $x_a,a=4,5,6,7$ breaks the $SO(4)=SU(2)_L \times SU(2)_R$ down to $SU(2)_L$.

The relevant parts of the D7 brane action are given by
\be\label{borninfeld}
S_{D7}= -N_f \mu_7 \int\! d^8x \, \sqrt{- \det(g_{ab}+ F_{ab})}+ {1\over 2} N_f \mu_7 \int P[C_{(4)}] \wedge F\wedge F,
\ee
where $C_{(4)}$ is given by
\be
C_4 = {(r^2+X_8^2 + X_9^2)^2} dt\wedge dx_1\wedge dx_2\wedge dx_3 + \dots
\ee
The $\dots$ represent terms in $C_{(4)}$ which do not have legs along $t,x_1,x_2,x_3$ but have to be present in order to guarantee the self-duality condition $F_{(5)} = \star F_{(5)}$ for the field strength $F_{(5)} = d C_{(4)}.$

For the gauge field ansatz (\ref{gaugeansatz}) the Wess Zumino term of the D7 brane action reads
\be\label{WZac2}
S_{WZ} = - 2 \mu_7 \, \mbox{vol}(S^3) \int\! d^4x dr \, (r^2 + X_8^2)^2 f \partial_r f.
\ee
The DBI term of the D7 brane is not illuminating and will be evaluated later in (\ref{detfac}) and in (\ref{ddefine}). 

Let us first study study the effects of the gauge field ansatz (\ref{gaugeansatz}) and in particular determine the conformal dimension of the dual operator. Therefore we study  the linearized equation of the gauge field for a D7 brane located at $X_8=L, X_9=0.$ With the ansatz (\ref{gaugeansatz}) the term  quadratic in $f$  of the action for the Born-Infeld part is given by (converting to polar coordinates which introduces a $r^3$ term in the integration measure)

\be
S_{DBI, quad} = - N_f \mu_7 \, \mbox{vol}(S^3) \int\! d^4x dr \,  \Big\{{2\over r} (L^2+r^2)^2 f^2+{1\over 2} r (\partial_\mu f)^2 +{1\over 2} r (L^2+ r^2)^2 (\partial_r f)^2\Big\},
\ee
while the WZ term (\ref{WZac2}) gives
\bea
S_{WZ} &=& -N_f \mu_7 \, \mbox{vol}(S^3) \int\! d^4x dr \, 2 (L^2+r^2)^2 f \partial_r f \no\\
&=&  -N_f \mu_7 \, \mbox{vol}(S^3) \int\! d^4x dr \, (L^2+r^2)^2 \partial_r (f^2) \no\\
&=&  4 N_f \mu_7 \, \mbox{vol}(S^3) \int\! d^4x dr \,  r (L^2+r^2) f^2.
\eea
The equation of motion of $f$ for the action $S_{DBI,quad} + S_{WZ}$ becomes
\be
{1\over r} \partial_r \big( r (r^2+L^2)^2 \partial_r f\big) +(\partial_\mu f)^2 - 4 {(r^2+L^2)^2\over r^2} f +8  (r^2+L^2) f=0.
\ee
Near the AdS boundary the asymptotic behavior of f is given by
\be\label{sourcef}
\lim_{r\to \infty} f(r) = c_+ (x){1\over r^{2}} + c_-(x){ \log r\over r^2} + \mathcal{O}\left((r^{-4}\right). 
\ee 
By the standard holographic dictionary, turning on $c_+ (x)$ corresponds to turning on a source for a operator of dimension $\Delta=2$. 

We conclude that the gauge field ansatz (\ref{gaugeansatz}) transforms as $(0,1)$ under $SU(2)_L \times SU(2)_{\cal R}$ and has conformal dimension $\Delta =2.$ Since it is the only type of fluctuation with these properties, $c_+(x)$ as defined by equation (\ref{sourcef}) sources the counterterm (\ref{counter3}).

\subsection{Equations of motion}
The   gauge field (\ref{gaugeansatz}) is clearly not the most general form and one has to check that our ansatz  is consistent. This means that the equations of motion for the gauge fields allow for all other components to be set to zero consistently. This is equivalent to the statement that all components of 
\be\label{eofmf0}
\int d^8x \Big\{\partial_a \Big( \sqrt{ -det(G+F)} \big(G+F\big)^{[ab]} \Big)  + \Big(\partial_a (r^2 +X_8^2)^2  \Big) \epsilon^{acdb}\partial_c A_d\Big\} \delta A^b =0
\ee 
can be reduced to a single partial differential equation for $f(r,x_3)$. One can indeed show that the $b=3$ component of  the bracket in (\ref{eofmf0}) is identically satisfied, whereas the $a=4,5,6,7$ components  are given by
\be
{\cal F}\Big(f,X_8, \partial f, \partial X_8 , \partial^2 f ,\partial^2 X_8) y^a=0,
\ee
 where $y^a$ is defined in (\ref{gaugef}).
The function ${\cal F}$ is a scalar function which is  invariant under $SO(4)$ rotations in the  $x_4,x_5,x_6,x_7$  space. Furthermore it depends only on $f,X_8$ and their derivatives with respect to $r$ and $x_3$. If one expands the variation $\delta A^a$ in terms of spherical harmonics, all of them vanish but the one proportional  to (\ref{gaugef}). This implies that there is no linear coupling of   (\ref{gaugef}) to other spherical harmonics in the action\footnote{Such a statement can also be derived from purely group theoretical arguments.} and hence  the truncation is consistent.

Both the equation of motion for $X_8$ and $f$ are highly nonlinear coupled second order partial differential equations whose form is very complicated and not very illuminating. We will not display their explicit form in this paper but they can be easily derived from the action (\ref{borninfeld}).

\section{$\kappa$-symmetry for the D7 brane probe}
\setcounter{equation}{0}
\label{sec3}

The $\kappa$-symmetry transformation on the world volume of the D7 brane  takes the form
\be
\delta \Theta= (1+ \Gamma) \kappa
\ee
and is responsible for  gauging away half the spinor degrees of freedom ensuring space time supersymmetry.
The condition for an unbroken supersymmetry for the D7 brane probe is given by
\be\label{kappaproj}
(1-\Gamma)\epsilon=0,
\ee
where $\epsilon$ are Killing spinors generating the  32 supersymmetries in the $AdS_5\times S^5$ background  \cite{Bergshoeff:1997kr}. It was shown in   
  \cite{Kehagias:1998gn} (see also \cite{Grana:2000jj}) that  for the $AdS_5\times S^5$ space parameterized by (\ref{adsmeta})   16 of the supersymmetries  the AdS Killing spinors can be written   in terms of  constant spinor
 
 \be\label{killingsp}
    \epsilon = {(r^2+\rho^2)}^{-{1\over 4}} \epsilon_0 ,
  \ee
  where $\epsilon$ is a doublet of ten dimensional  Majorana Weyl spinors which satisfy
  \be\label{mjweyl}
  i\sigma_2 \otimes \Gamma_{0123}\;  \epsilon_0 = \epsilon_0.
  \ee
  The spinor  (\ref{killingsp})   generates   the sixteen Poincare supersymmetries. In addition there are sixteen  superconformal supersymmetries. For an embedding with  nonzero $X_8$  conformal invariance is broken and we only need to consider the Poincare supersymmetries. Since the $r$ dependence of (\ref{killingsp}) is an overall factor, we can drop the overall factor in   (\ref{killingsp}) and work with a constant spinor instead.

The form of the $\kappa$-symmetry projector for the D-brane probe  is given by \cite{Bergshoeff:1997kr}.
\be\label{gammapro}
\Gamma=  {1\over \sqrt{-\det(G+F)}} \sum_{n=0}^\infty {1\over 2^n n!}\gamma^{i_1 j_1 i_2 j_2 \cdots i_n j_n} F_{i_1 j_1} F_{i_2 j_2} \cdots F_{i_nj_n} J_{7}^{(n)},
\ee
where 
\be
J_7^{(n)} = (-1)^n  (\sigma_3)^n  \; i \sigma_2\otimes \gamma_{01234567}.
\ee
The gamma matrices $\gamma_i$ are the pull-backs of the tangent space gamma matrices  on the world volume and given by
\be\label{pullbackgam}
\gamma_i = E^A_\mu \partial_i X^\mu \Gamma_A.
\ee
The explicit expressions for $\gamma_i$ can be found in appendix \ref{appendc}.
For the case of the D7 brane and a gauge field ansatz (\ref{gaugeansatz}) the expansion of the projector  in the field strength terminates at quadratic order.  The $\kappa$-symmetry projector (\ref{gammapro}) can be expressed in terms of three prices $\Gamma_{n=k},\;  k=0,1,2$ with zero, one and two powers of the gauge fields respectively.
\bea
\Gamma&=& \Gamma_{n=0}+ \Gamma_{n=1}+\Gamma_{n=2}\no\\
&=&  {1\over \sqrt{-\det(G+F)}}  \Big[ i \sigma_2\otimes \gamma_{01234567}
 - \big(  \gamma^{3a} F_{3a}+{1\over 2}\gamma^{ab}F_{ab}\big) \sigma_1\otimes \gamma_{01234567}\no\\
 && \qquad \quad \qquad\qquad + \left( {1\over 2} \gamma^{3abc} F_{3a} F_{bc} + {1\over8} \gamma^{abcd} F_{ab} F_{cd} \right) i \sigma_2 \otimes  \gamma_{01234567}\Big].
\eea
The determinant $-\det(G+F)$ contains the metric determinant of the three sphere which can be extracted, where we define
\be\label{detfac}
-\det(G+F) = D
\ee
and $D$ is given by
 \bea\label{ddefine}
D&=& {\Big(r^4+ 4 f^2 (r^2+X_8^2)^2\Big) \over r^6} \left\{  \Big( {r^2\over (r^2+X_8^2)^2}+  (\partial_r f)^2 
\Big) (\partial_{3} X_8)^2 + \Big( r^2+ (\partial_3 f)^2 
\Big) (\partial_{r} X_8)^2\right.\no\\
&&   -  2 \partial_3 f \partial_r f 
 \partial_{r} X_8 \partial_{3} X_8 +  \Big( r^2+  (\partial_3 f)^2 + (r^2+X_8^2)^2 (\partial_r f)^2\Big)
\bigg\}.
\eea

\subsection{Projection conditions on the spinors}
In order to  simplify the form of the $\kappa$-symmetry projector and solve the conditions for the existence of 
unbroken supersymmetries we will impose three projection conditions on the infinitesimal supersymmetry transformation parameter $\epsilon$. 
\bea
i\sigma_2 \otimes \Gamma_{0123} \epsilon&=&\epsilon,  \label{proja}\\
1_2\otimes\Gamma_{4567}  \epsilon&=&\epsilon, \label{projb}\\
\sigma_3\otimes \Gamma_{38} \Gamma_{45}\epsilon \label{projc} &=& \epsilon.
\eea
Note that the first projection (\ref{proja}) was already imposed by singling out the Poincare supersymmetries (\ref{mjweyl}).

Since the three matrices on the left hand side all commute,  the three projectors are compatible and reduce the 32  supersymmetries  of the $AdS_5\times S^5$ background to 4.  The  expressions for $\Gamma_{n=0,1,2}$  are evaluated in detail in appendix  \ref{appendb}, where the final form of the projector $\Gamma= \Gamma_{n=0}+\Gamma_{n=1}+\Gamma_{n=2}$  can be found in  (\ref{gammafullappend}). We display this result again for the convenience of the reader:
\bea\label{gammafull}
\Gamma \, \epsilon &=& {1\over \sqrt{D}} \Big\{ 1+ {\partial_r X_8 \over r}\partial_3 f -{2\over r^3} (r^2+X_8^2)^2 f \partial_r f - \partial_3 X_8 \big({2f\over r^2}+ {\partial _r f \over r}\big) \Big\} 1_2\otimes 1\, \epsilon \no \\
&+& {(r^2+X_8^2)\over \sqrt{D}} \Big\{  {-\partial_3 X_8\over (r^2+X_8^2)^2} -  \big({2f\over r^2}+ {\partial _r f \over r}\big) +{2 f\over r^3}  \big( \partial_r f \partial_3 X_8 - \partial_3 f \partial_r X_8\big)\Big\} 1_2\otimes \Gamma_{38} \, \epsilon\no\\
&+&  {2(r^2+X_8^2) f \over r^3  \sqrt{D}} \Big\{ {\partial_3 f\over r}-\partial_r X_8\Big\} 1_2\otimes x^d \Gamma_{d3} \, \epsilon  +{1\over \sqrt{D}}{1\over r}\Big\{ {\partial_3 f\over r}-\partial_r X_8\Big\} 1_2\otimes x^d \Gamma_{d8} \, \epsilon. \no \\ 
\eea
It is remarkable after employing the conditions (\ref{proja})-(\ref{projc}) the complete projector can be written as the sum of four terms involving only four linearly independent combinations of gamma matrices. The conditions  (\ref{proja})-(\ref{projc})  reduce the Poincare supersymmetry from 32 to 4. Consequently the $\kappa$-symmetry projector (\ref{kappaproj}) should not reduce the number of unbroken supersymmetries further.

It follows that $\Gamma$  has to be equal to the unit operator. This implies  that   the term in (\ref{gammafull}) proportional to  $1_2\otimes 1$ has to be equal to one, or equivalently
\be
1+ {\partial_r X_8 \over r}\partial_3 f -{2\over r^3} (r^2+X_8^2)^2 f \partial_r f - \partial_3 X_8 \big({2f\over r^2}+ {\partial _r f \over r}\big) =\sqrt{D}, \label{bpseqc}
\ee
where $D$ is given by (\ref{ddefine}). 
Furthermore, the linearly independent terms proportional to $1_2\otimes \Gamma_{38}$, $1_2\otimes x^d \Gamma_{d3} $ and $1_2\otimes x^d \Gamma_{d8}  $ have to vanish separately,  Note that the vanishing of  terms  proportional to $1_2\otimes x^d \Gamma_{d3}$ and $1_2\otimes x^d \Gamma_{d8}  $    give the same condition, hence we have two additional BPS equations

\bea
{\partial_3 f\over r}-\partial_r X_8&=& 0, \label{bpseqa} \\
-{\partial_3 X_8\over (r^2+X_8^2)^2} -  \big({2f\over r^2}+ {\partial _r f \over r}\big) +{2 f\over r^3}  \big( \partial_r f \partial_3 X_8 - \partial_3 f \partial_r X_8\big)&=&0 .\label{bpseqb} 
\eea
 A lengthy calculation shows that  if (\ref{bpseqa}) and (\ref{bpseqb}) are satisfied then both  (\ref{bpseqc}),  as well as the equations of motion, are   automatically satisfied. Hence finding a D7 brane embedding which preserves four supersymmetries   boils down to finding solutions two BPS equations (\ref{bpseqa}) and (\ref{bpseqb}).

\section{Solutions of the BPS equations}
\setcounter{equation}{0}
\label{sec4}

The main result from the last section was the the there are four  unbroken supersymmetries if the BPS equations  (\ref{bpseqa}) and (\ref{bpseqb}) are satisfied. The two equations are considerably simpler than the equations of motion, since they involve only first derivatives. 
They are however still nonlinear partial differential equations and hence not easy to solve. In the following we will present some solutions to the BPS equations  using a variety of methods. The solution corresponding to a supersymmetric holographic topological insulator will be obtained numerically.

\subsection{A solution using scaling symmetry}\label{scalesol}

The two BPS equations  (\ref{bpseqa}) and (\ref{bpseqb}) are invariant under the following scaling transformation (which generalizes the scaling transformation of \cite{HoyosBadajoz:2010ac}).

\be
r\to \xi r, \quad x_3 \to {1\over \xi} x_3, \quad X_8 \to \xi X_8, \quad f \to  f.
\ee
One can use the scaling symmetry to obtain an ansatz for the fields which only depend on the scaling invariant combination $y= x_3 r$

\be
X_8 = r  \;  h( y), \quad  f = f(y) .
\ee
The BPS equations become
\bea\label{bpsc}
y \; h' + h -f'&=&0,\nonumber\\
{h'\over (1+h^2)^2} +2 f + (y+2 f \,h)f' &=&0.
\eea
The first equation can be integrated  and determines $h$ 
\be
h = {c_1\over y} + {f\over y}.
\ee
The second equation of (\ref{bpsc}) then takes the following form
\be
f'= -\frac{y  \left( -{c_1 \over y^2} -{f\over y^2} + 2 f \Big( 1+ {(c_1+f)^2\over y^2}\Big)^2 \right)}{1+ \Big(y^2+ 2 f (c_1+f) \Big)\Big( 1+ {(c_1+f)^2\over y^2}\Big)^2  }.
\ee

\begin{figure} 
\begin{center}
\includegraphics[scale=0.30]{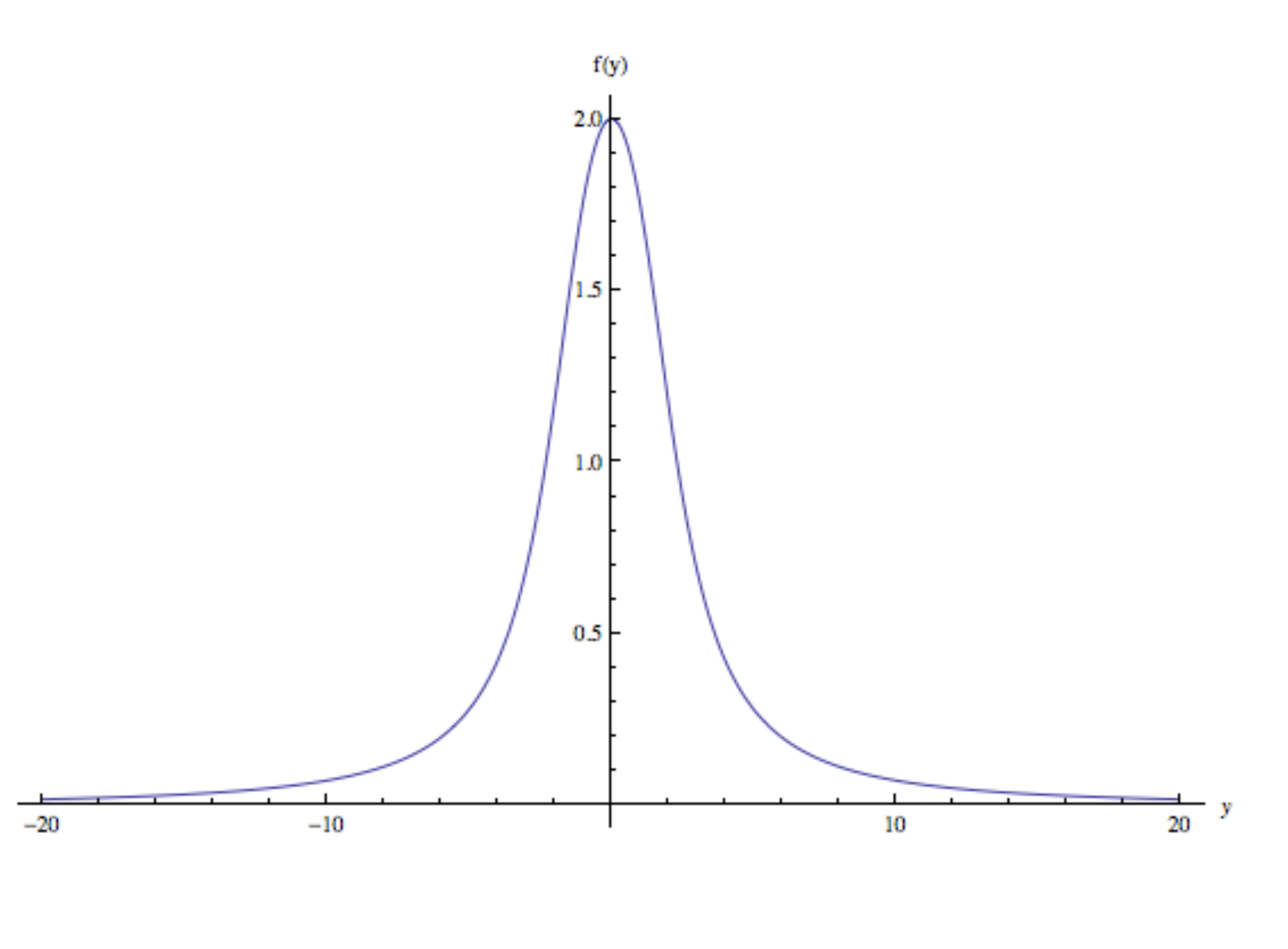}\includegraphics[scale=0.36]{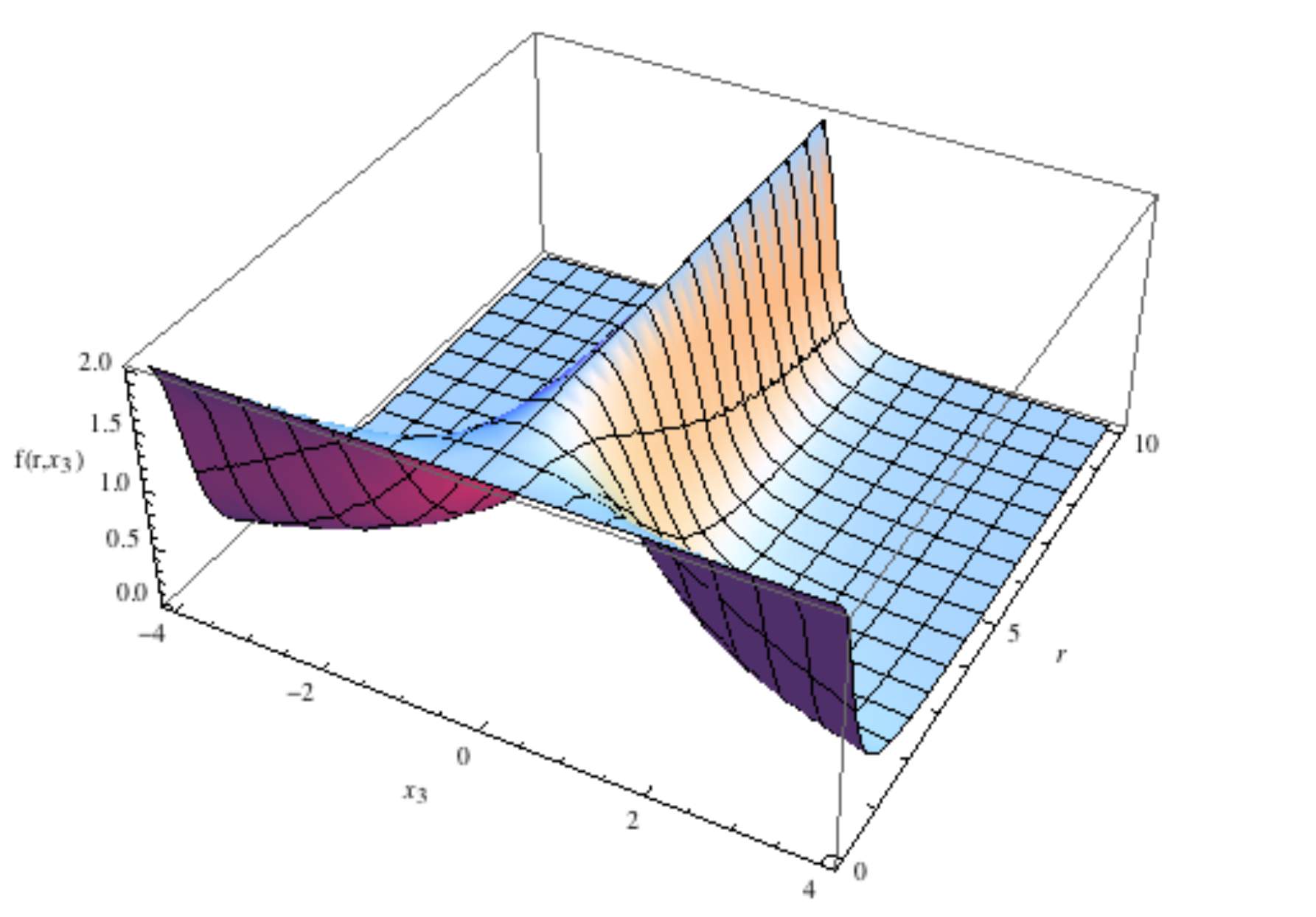}\includegraphics[scale=0.33]{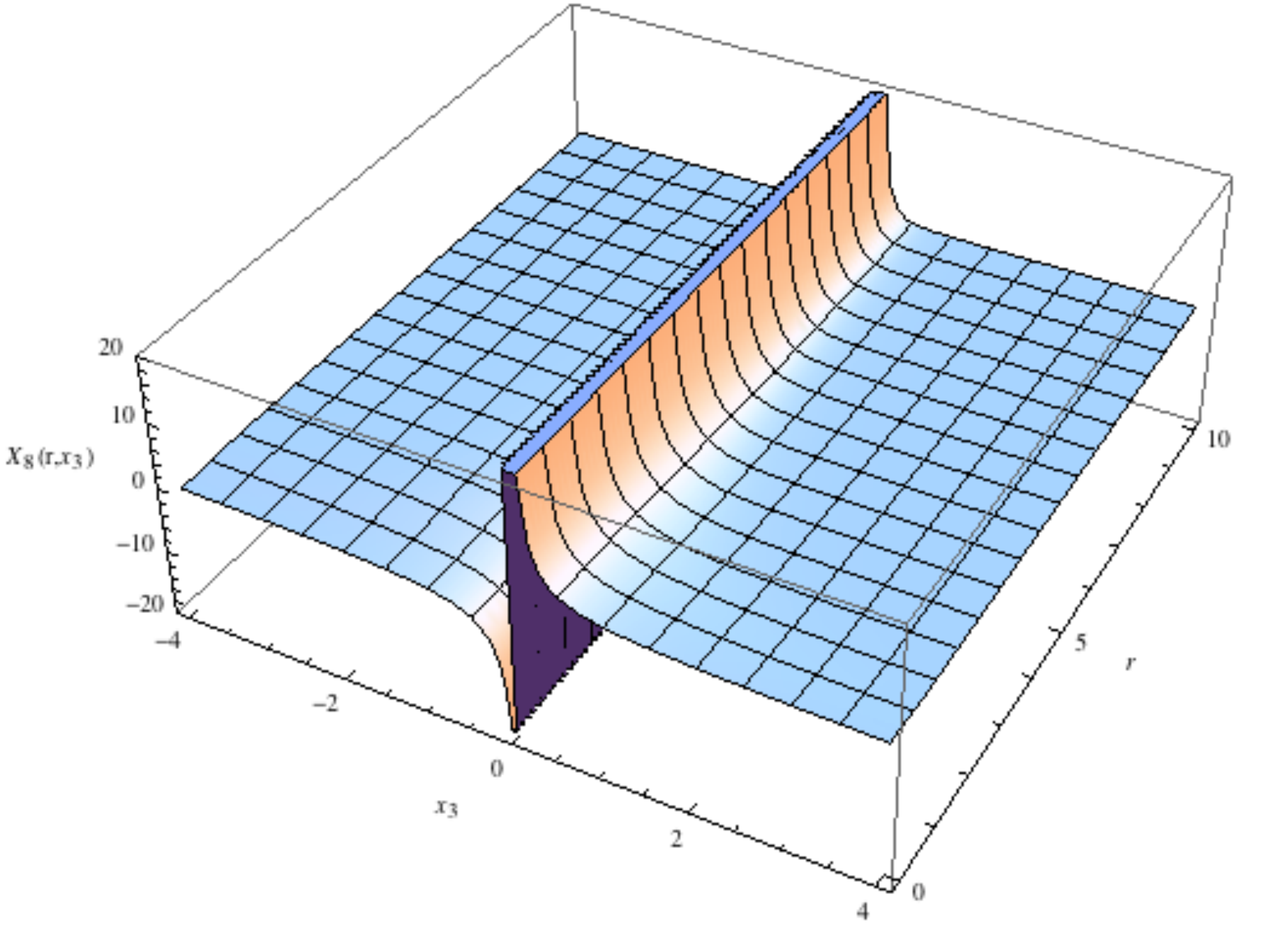}
\caption{(a) numerical solution for $f(y)$,  (b) Plot of $f(r,x_3)$,  (c) Plot of $X_8(r,x_3)$}
\label{fig2}
\end{center}
\end{figure}

While this ordinary differential equation does not seem to have an explicit solution in terms of known functions, it can be easily integrated numerically. Note that if one imposes the initial condition $f(y)|_{y=0} \neq 0$ the  solution indeed satisfies  $\partial_r X_8|_{r=0}=0$ which is equivalent to the condition that the 3-sphere closes off smoothly at $r=0$.  In figure \ref{fig2} we display an example for $c_1=0$ and  $f(y)|_{y=0}=2$.

Note that this solution does not correspond  to a topological insulator. This follows from the fact that  $\lim_{x_3\to \pm \infty}  X_8 =0$ which means that the fermion mass cannot behave as in (\ref{limitx8}).

\subsection{Perturbation theory}

In this section we will give a perturbative solutions of the
 BPS equations (\ref{bpseqa}) and (\ref{bpseqb}). First, we redefine $f = r \tilde f$ such that the BPS equation (\ref{bpseqa})  becomes
\bea\label{newa}
\partial_3 \tilde f = \partial_r X_8
\eea
while the second BPS equation (\ref{bpseqb}) reads
\bea\label{newb}
\partial_r \tilde f + {3\over r} \tilde f + {\partial_3 X_8 \over (r^2 + X_8^2)^2} + {2 \tilde f \over r^2} \Big( r \partial_{3} \tilde f \partial_r X_8 - \partial_3 X_8(\tilde f + r \partial_r \tilde f)\Big)=0.
\eea
Equation (\ref{newa}) can be solved by
\be\label{newc}
\tilde f = \partial_r h + \partial_3 \tilde h , \quad X_8 = \partial_3 h - \partial_r \tilde h,
\ee
where $\tilde h$ can is a harmonic function satisfying 
\be \label{sol}
\left(\partial_3^2 + \partial_r^2 \right) \tilde h =0.
\ee
We should consider $\tilde{h}$ as a   fluctuation on top of the embedding. Since we are only interested in the embedding, we set $\tilde{h}=0$ from now on and plug equation (\ref{sol}) into the second BPS equation (\ref{newb}). This leads to a complicated partial differential equation for the function $h$
\bea\label{newd}
{\partial_3^2 h \over (r^2 + (\partial_3h)^2)^2}+ {3 \partial_r h\over r} + {2 \partial_3^2 h \; (\partial_r h )^2 \over r^2}+ {2 \partial_r h (\partial_r \partial_3 h)^2\over r} - {2\partial_3^2 h \;\partial_r^2 h\; \partial_r h \over r} + \partial_r^2 h =0.
\eea
Unfortunately, we were not able to solve equation (\ref{newd}) analytically. Following \cite{HoyosBadajoz:2010ac}   we take  the ansatz for the perturbation expansion for $h$ as follows
\be\label{pertuba}
h= \epsilon {1\over r} h_1( r x_3) +  \epsilon^{3} {1\over r^3} h_3( r x_3)+\cdots= \sum_{n=1} \left({ \epsilon\over r}\right)^{2n-1} h_n(r x_3) .
\ee
The order $\epsilon$ contribution to (\ref{newd}) is given by, renaming $r x_3 =y$
\be
h_1(y)- y h_1'(y)- (1+y^2) h_1''(y)=0,
\ee
which has the most general solution 
\be\label{1stordersol}
h_1(y)= M_0(1+y^2)^{1/2} + c_1 y.
\ee
In the following we set $c_1=0$ since it corresponds to a constant mass term.
In this case the order $\epsilon^3$ contribution to (\ref{newd}) has the form
\be\label{thirdord}
h_2''(y) - {3 y\over1+y^2} h_2' + {3\over 1+y^2} h_2 = 2M_0^3  {(y^2-1)\over (1+y^2)^{7/2}}.
\ee
This equation can be integrated and one gets
\be
h_2(y)= - M_0^3 { 1 \over 6 (1+y^2)^{3/2} } \Big( 1+  6y^2+4y^4\Big) + c_{2,1} y + c_{2,2} \Big( 3y \; {\rm arcsinh(y)} + \sqrt{1+y^2}( y^2-2)\Big) ,
\ee
where the second and third terms are solutions to the homogeneous part of the (\ref{thirdord}). We note that the term of order $\epsilon^1$ is dominant near the boundary, i.e. as $r\to \infty$ one obtains from (\ref{1stordersol}) and (\ref{newc})
\be
\lim_{r\to \infty} X_8 = \lim_{r\to \infty} M_0{x_3 r \over \sqrt{1+ (x_3 r)^2}} + \mathcal{O}\left({1\over r^2}\right) = M_0 {x_3\over |x_3|} + \mathcal{O}\left({1\over r^2}\right) .
\ee 

It is in principle possible to find the  higher orders of the perturbation series (\ref{pertuba}), which are all sub-leading in powers of $r$. However the perturbative solution shares the same problem as \cite{HoyosBadajoz:2010ac}.  For the leading term in $\epsilon$ one has 
\be
\lim _{r\to 0} \partial_r X_8 = M_0 x_3+\cdots
\ee
One can easily convince oneself that the non vanishing of $\partial_r X_8$ at $r=0$ is a general feature 
of the perturbation series (stemming from the scaling symmetry of section \ref{scalesol}). 
However  $\partial_r X_8$  has to vanish at $r=0$ for the sphere to close off smoothly. Consequently,   while the perturbation expansion  has a reasonable behavior at $r\to \infty$,  the behavior as $r\to 0$ does not correspond to a smooth solution.

\subsection{Numerical solution of the BPS equations}
In  this section we solve the BPS equation (\ref{newd}) numerically by using a heat method.  This method was also used to obtain solutions for the non-supersymmetric setup of \cite{HoyosBadajoz:2010ac}.

\begin{figure} 
\begin{center}
\includegraphics[scale=0.28]{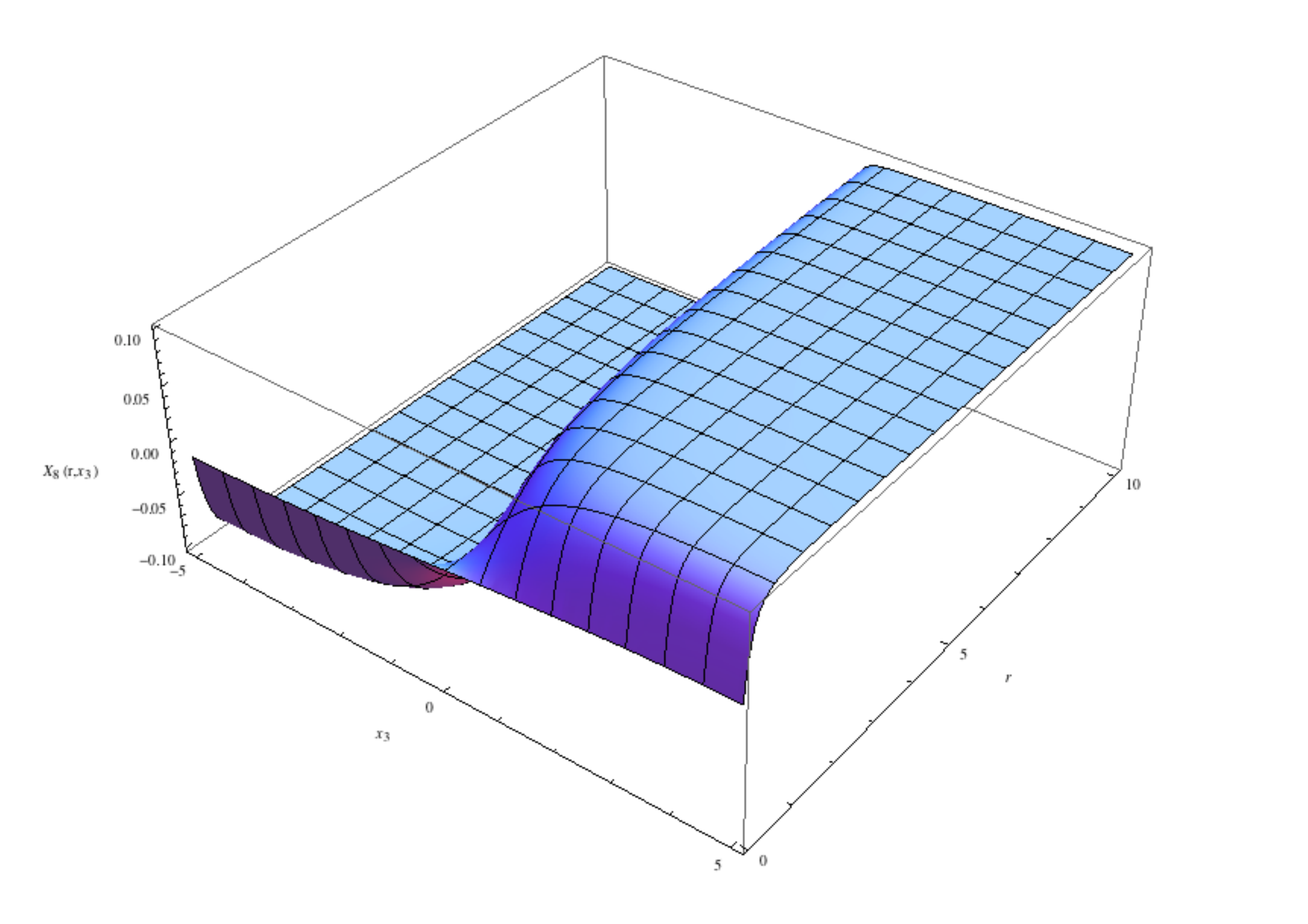}\includegraphics[scale=0.28]{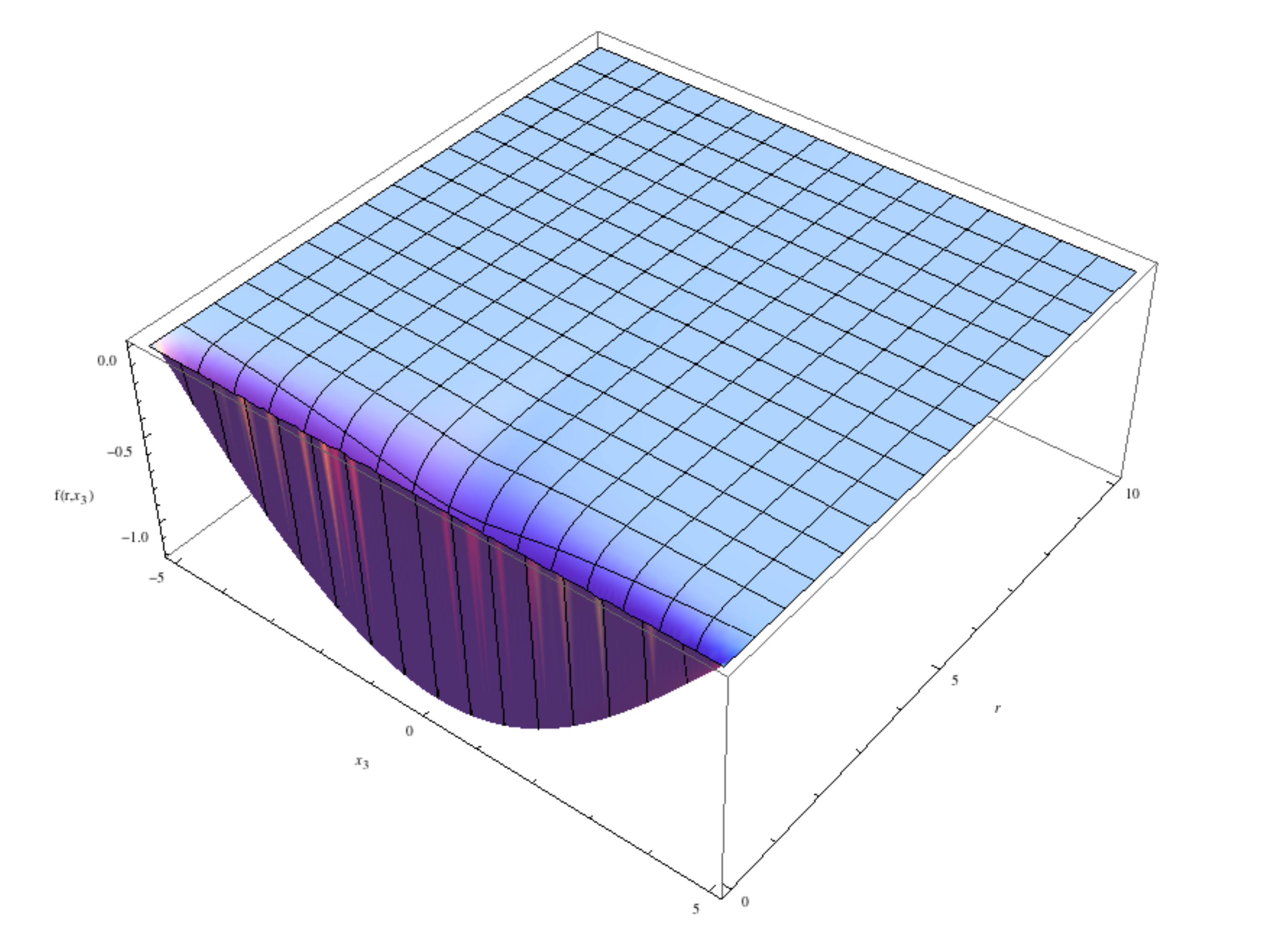}
\caption{(a) numerical solution for $X_8(r,x_3)$,  (b) Plot of $f(r,x_3)$} for large $\tau$.
\end{center}
\label{fig3}
\end{figure}

In particular we make $h(r,x_3)$ dependent on  an imaginary time $\tau$.  The derivative with respect to $\tau$ is given by the left-hand side of equation (\ref{newd}) multiplied by $r^2$.
\bea
{\partial  h(r,x_3,\tau)\over \partial\tau}&=&
{r^2 \partial_3^2 h \over (r^2 + (\partial_3h)^2)^2}+ r^2\partial_r^2 h + {3r\;  \partial_r h} + {2 \partial_3^2 h \; (\partial_r h )^2 }\nonumber\\
&&\quad + {2 r \;\partial_r h (\partial_r \partial_3 h)^2} - {2 r \;\partial_3^2 h \;\partial_r^2 h\; \partial_r h } .
\eea

After a sufficiently long time $\tau$ the field configuration approaches a solution of the right hand side equation.
We solved this equation for a spatial region $x_3=[-x_{max},x_{max}]$ and $r\in [0,r_{max}]$.  As a good initial function $h$  at time $\tau=0$ we use the first order solution (\ref{1stordersol}) (with $c_1=0$) which we can trust for large $r$. At the boundaries of the spatial regions, we impose Neumann boundary conditions for $h$ at $r=0$ and Dirichlet boundary conditions (given by the first order solution).   The numerical  solution of the equations was obtained using Mathematica and unlike the perturbative solution the numerical solution has the correct behavior both in the $r=0$ region and in the large $r$ region.

\section{Discussion}
\setcounter{equation}{0}
\label{sec5}

Making coupling constants dependent on space time coordinates generically breaks all the supersymmetries of a supersymmetric field theory. Examples are Janus solutions \cite{Bak:2003jk} and the holographic topological insulator solutions constructed in \cite{HoyosBadajoz:2010ac}. Turning on a world volume gauge field allowed us to preserve four of the sixteen supersymmetries of the probe D7 brane.  On the field theory side this corresponds to  turning on localized counterterms as expected from the weak coupling analysis. A supersymmetric generalization of the topological insulator might be interesting in its own right from a theoretical perspective.

The BPS equations we have found are considerably simpler than the equations of motion, since they are first order instead of second order partial differential equations. Utilizing the scaling symmetry we have been able to generate exact smooth solutions. These solutions do however not have the boundary behavior which describes topological insulators. We have been unable so far to obtain closed form solutions corresponding to holographic topological insulators.  
Instead we had to resort to other methods such as   perturbation theory and  numerical methods.  A closed form solution would be very useful as it is possible to study fluctuations about the solution and calculate correlation functions and transport quantities.

 It is an interesting open problem to determine whether the difficulty of obtaining closed form solutions  is a fundamental problem due to the small amount of preserved supersymmetry or simply due to the lack of ingenuity on the authors' side\footnote{Note that in the case of supersymmetric Janus solutions \cite{D'Hoker:2007xy}
 the original form of the BPS equations  looked too complicated to be solved. Only a sequence of clever variable changes made a solution possible.}.

Even with the numerical solutions it might be feasible to extract observable quantities, like transport quantities from the numerical solutions. We leave such questions for future work.

\newpage

\noindent{\Large \bf Acknowledgements}

\bigskip

This  work was in part supported  by NSF grant PHY-07-57702. We are grateful to Carlos Hoyos, Kristan Jensen, Andreas Karch and Andy O'Bannon for useful conversations and correspondence.  We are grateful to Brian Shieh for initial collaboration on this project.  MG gratefully acknowledges the hospitality of the Newton Institute for Mathematical Sciences while this paper was finalized.

\appendix

\section{Conventions}
\setcounter{equation}{0}
\label{appenda}

We will briefly review our  conventions for the IIB supergravity  and ten dimensional gamma matrices  which follow \cite{Bergshoeff:1997kr}.  
The supersymmetry transformation parameters $\epsilon$ of  type IIB supergravity can be written as a doublet
\be\label{doubletspin}
\epsilon= 
\left(
\begin{array}{c}
  \epsilon_1   \\
      \epsilon_2
\end{array}
\right).
\ee
The flat space  gamma matrices $\Gamma_i$ are $32\times 32$ matrices and the spinors $\epsilon_i,i=1,2$ are Majorana-Weyl spinors
\be
\Gamma_{11} \epsilon_i = \epsilon_i, \quad i=1,2.
\ee
The matrices in the $\kappa$-symmetry projectors are of the form
\be
\sigma_i \otimes \Gamma_{i_1 i_2 \dots i_n} ,
\ee
where $\Gamma_{i_1 i_2 \cdot i_n} = \Gamma_{i_1} \Gamma_{i_2} \dots \Gamma_{i_n}$ and the Pauli matrices $\sigma_i$ act on the doublet  (\ref{doubletspin})
 
\section{Supersymmetry and  counter-terms}
\setcounter{equation}{0}
\label{appendd}

 In this appendix we   review  and adapt the results given in \cite{Clark:2004sb}  for the field theory description of the topological insulator discussed in the body of the paper.
 
 The  theory is most conveniently formulated  in terms  in $N=1$ chiral multiplets, where only the $N=1$ supersymmetry is manifest.  We consider $n$ chiral multiplets  denotes $\Phi^i$ with the field content.
\be
\Phi^i:  \; \phi^i, \; \psi^i, \; F^i,\quad i=1,2,\cdots, n,
\ee
where $\phi^i$ is a complex scalar, $\psi^i$  is a Majorana spinor and $F^i$ is an auxiliary field.
In components the Lagrangian density  is given by
\bea\label{lagrang}
{\cal L}&=& - \partial_\mu \phi^i \partial^\mu \phi^{i*}-{i\over 2} \bar \psi^i \gamma^\mu \partial_\mu \psi^i + F^{i*} F^i  -{i\over 2} {\partial W\over \partial \phi_i \partial \phi_j} \bar \psi^i P_+ \psi^j + {\partial W\over \partial \phi_i} F^i + c.c,
\eea
where $P_+ ={1\over 2}(1+\gamma^5)$. The supersymmetry transformations are given by
\bea\label{susytraf}
\delta \phi^i &=& i \sqrt{2} \bar \psi^i P_+ \epsilon \no\\
\delta (P_+ \psi^i) &=& \sqrt{2} P_+ \epsilon\;  F^i +\sqrt{2} \gamma^\mu P_- \epsilon\;  \partial_\mu \phi^i \no\\
 \delta F^i &=& - i \sqrt{2} \bar \epsilon \gamma^\mu \partial_\mu P_+ \psi^i.
\eea
Here  $\epsilon$ is a Majorana spinor parameter.  $W$ is the superpotential which parameterizes the self interactions of the chiral superfields.  It was shown  \cite{Clark:2004sb}  that couplings  in the superpotential which position dependent  break all of the  supersymmetries  (\ref{susytraf}). Assuming that the superpotential depends on one coupling constant $g(x_3)$, the supersymmetry variation of the  Lagrangian (\ref{lagrang})   is given by
\be
\delta {\cal L}=  i \sqrt{2}{\partial g\over \partial x_3} \; \sum_i  \bar \epsilon \left( P_+ \gamma^3 \psi_i   {\partial \over \partial g} \left({\partial W\over \partial \phi_i}\right)^* +P_- \gamma^3 \psi_i   {\partial \over \partial g} \left({\partial W\over \partial \phi_i}\right)\right)
\ee
is not a total derivative. Consequently,  all supersymmetries are broken by the position dependent mass. Furthermore it was shown in \cite{Clark:2004sb} that some - but not all - supersymmmetries can be restored by adding a counter-term  to the action of the form
\be\label{counter}
{\cal L}\to {\cal L}+ i   {\partial g\over \partial x_3} \left\{ \left({\partial W\over \partial  g}\right) - \left({\partial W\over\partial g}\right)^*\right\}.
\ee
The  supersymmetry which is preserved is given by
 \be\label{susyproj}
\Pi \epsilon=\epsilon, \quad \quad \Pi ={1\over 2} \big( 1+ i \gamma^5 \gamma^3\big) .
\ee
For the theory discussed in section \ref{ftanalys} the coupling $g$ is identified with the mass $m$, the chiral superfields $\Phi_1, \Phi_2$ are identified with $Q, \tilde Q$ the superpotantial is given by
\be
W= m(x_3) \,Q \tilde Q
\ee

\section{D7 brane embedding}
\setcounter{equation}{0}
\label{appendc}
 
In this appendix we present the details of the D7 brane embedding for completeness. The $AdS_5\times S^5$ background metric is given by
\be\label{metriccar}
ds^2 = g_{\mu\nu} dx^\mu dx^\nu= (r^2+\rho^2) \eta_{ij} dx_i dx_j +  \frac{1}{r^2 +\rho^2} \left( \sum_{a=4}^7 dx_a^2 + dX_8^2 + dX_9^2 \right),
\ee
where
\be 
r^2=x_4^2+x_5^2+x_6^2+x_7^2, \quad \quad \rho^2= X_8^2+ X_9^2.
\ee
Note that we have introduced two different set of indices: $i, j \in \{ 0, 1,2,3 \}$ and $a \in \{ 4, 5, 6, 7 \}.$ The corresponding coordinates $x_i$ and $x_a$ parametrize the field theory coordinates and the internal coordinates, respectively.
The D7 brane world volume coordinates $\zeta$ are identified with $x_i (i=0,1,2,3)$ and $x_a (a=4,5,6,7).$ In order to embed the D7 brane into $AdS_5\times S^5$ we have to specify $X_8$ and $X_9$ as a function of the world volume coordinates. In this paper we restrict ourselves to 
\bea\label{embeddingcar}
X_8 &=& X_8(x_3, r),\no\\
X_9 &=& 0 \, .
\eea
The induced metric on the world volume of the brane, with components $G_{ab},$ is defined by
\be
G_{ab}= \partial_a X^\mu \partial_b X^\nu g_{\mu\nu}
\ee
and takes the following form
\bea
G_{ij} &=& (r^2+X_8^2) \eta_{ij}, 
\no \\
G_{33} &=&  (r^2+X_8^2) + {1\over (r^2+X_8^2) } \partial_3 X_8 \partial_3 X_8,\no  \\
G_{3a} &=&  \frac{1}{r^2+X_8^2} \frac{x_a}{r} \partial_3 X_8 \partial_r X_8 
\no, \\
G_{ab}&=& {1\over r^2+X_8^2} \left(\delta_{ab}+(\partial_r X_8 )^2 {x_a x_b\over r^2 } \right) ,
\eea
where $i,j \in \{ 0, 1, 2 \}$ and $a,b \in \{ 4,5,6,7 \}.$
The non-vanishing components of the field strength are,
for $ a=4,5,6,7$ 
\bea\label{fieldstrength}
F_{3a} &=&{\partial_3 f(r,x_3) \over f(r,x_3) }A_a, \no\\
F_{45} &=&-\frac{2 f(r,x_3)}{r^2}-\frac{1}{r} \partial_r \left( \frac{f(r,x_3)}{r^2} \right)(x_4^2+x_5^2),\no\\
F_{46} &=&-\frac{1}{r} \partial_r \left( \frac{f(r,x_3)}{r^2} \right)(x_4 x_7+x_5 x_6),\no\\
F_{47}&=& \frac{1}{r} \partial_r \left( \frac{f(r,x_3)}{r^2} \right)(x_4 x_6-x_5 x_7 ),\no\\
F_{56} &=& \frac{1}{r} \partial_r \left( \frac{f(r,x_3)}{r^2} \right)(x_4 x_6-x_5 x_7 ),\no\\
F_{57} &=& \frac{1}{r} \partial_r \left( \frac{f(r,x_3)}{r^2} \right)(x_4 x_7+x_5 x_6),\no\\
F_{67} &=&\frac{2 f(r,x_3)}{r^2} +\frac{1}{r} \partial_r \left( \frac{f(r,x_3)}{r^2} \right) (x_6^2+x_7^2).
\eea
Note that $F_{46} = - F_{57}$ and  $F_{47} = F_{56}.$ The pull back of the gamma matrices on the world volume (\ref{pullbackgam})  are given by
\bea\label{gammalowcar}
\gamma_i &=& \sqrt{ r^2+X_8^2} \; \Gamma_i , \quad i=0,1,2 ,\no\\
\gamma_3 &=&\sqrt{r^2+X_8^2} \;  \Gamma_3 \left( 1 + {\partial_3 X_8 \over r^2+X_8^2} \Gamma_{38} \right), \no\\
\gamma_a &=& {1 \over \sqrt{ r^2+X_8^2}} (\Gamma_a + {\partial_r X_8 \over r }x_a \Gamma_8) \quad a=4,5,6,7 . 
\eea

\section{Calculation of the $\kappa$-symmetry projector}
\setcounter{equation}{0}
\label{appendb}
In this appendix we will present the detailed calculation of $\Gamma_{n=0}, \Gamma_{n=1}$ and $\Gamma_{n=2}$. We use the three projection conditions (\ref{proja})-(\ref{projc}) as well as the properties of the world volume gauge field to bring the projector in a minimal form.
\subsection{Calculation of $\Gamma_{n=0}$ }
Using the expressions for the pulled back gamma matrices one get 
\bea
\Gamma_{n=0}&=& {1\over \sqrt{-\det(G+F)}}  i \sigma_2\otimes \gamma_{01234567}\nonumber\\
&=&{1\over \sqrt{D}} \; i \sigma_2\otimes \Gamma_{01234567}\left[ 1+{1\over (r^2+X_8^2)}\Gamma_{38} \partial_3 X_8 + {\partial_r X_8\over r } (x_a\Gamma_a) \Gamma_8 \right].
\eea
Employing the  projectors (\ref{proja}) and (\ref{projb})  $\Gamma_{n=0}$ can be expressed as follows. 
\be\label{bpszero}
\Gamma_{n=0} \, \epsilon ={1\over \sqrt{D}} \; \left[ {1_2\otimes 1}-{\partial_3 X_8 \over (r^2+X_8^2)}1_2\otimes \Gamma_{38} - {\partial_r X_8\over r } x^d 1_2\otimes\Gamma_{d8}\right] \epsilon.
\ee

\subsection{Calculation of $\Gamma_{n=1}$}
The contribution to $\Gamma$ which is linear in the field strength is given by
\bea\label{gammonea}
\Gamma_{n=1}&=& - {1\over \sqrt{D}}\left( \gamma^{3a}F_{3a} +{1\over 2} \gamma^{ab}F_{ab} \right) \sigma_1\otimes \gamma_{01234567}  \no\\
&=&  + {1\over \sqrt{D}}\left( {1\over 6} \epsilon^{abcd} F_{3a}\; \sigma_1\otimes\gamma_{012 bcd} + {1\over4}  \epsilon^{abcd} F_{ab} \;  \sigma_1\otimes\gamma_{0123 cd} \right),
\eea
where $\epsilon^{abcd}$ is the totally antisymmetric tensor in $(4567)$ space normalized such that $\epsilon^{4567}=1.$ 

We can use the formulae for the pull back of the gamma matrices (\ref{gammalowcar}) to determine 
\bea
\gamma_{0123cd}& =& (r^2+X_8^2) \Gamma_{0123cd} + \partial_3 X_8 \Gamma_{0128cd} +(r^2+X_8^2)  {\partial_r X_8\over r} \Gamma_{0123}(x_d \Gamma_{c8}-x_c \Gamma_{d8}) ,\no\\
\gamma_{012bcd}& =&  \Gamma_{012bcd} + {\partial_r X_8\over r} \left( x_b \Gamma_{012cd}\Gamma_8 - x_c \Gamma_{012bd} \Gamma_8+ x_d\Gamma_{012bc}\Gamma_8\right).
\eea
In addition we need the following relation involving the field strength
\be
\epsilon^{abcd} F_{ab} x_c \Gamma_d \;  \epsilon= -{4 f\over r^2} y_d \Gamma_d \; \epsilon,
\ee
where the coordinates $y_i$ were defined in (\ref{gaugeansatz}). There are three additional identities involving the field strength, which we need in the following
\bea
\epsilon^{abcd} F_{ab} \Gamma_{cd} \epsilon&=& \Big({1\over 2} \epsilon^{abcd} F_{ab}-F_{cd} \Big)\Gamma_{cd} \epsilon
= \left( 4 {f\over r^2}+ {2\partial_r f\over r} \right) \Big( \Gamma_{45}-\Gamma_{67}\Big)  \epsilon,
\eea
as well as 
\bea
\epsilon^{abcd} F_{3a}x_b \Gamma_{cd8} \epsilon= -\partial_3 f \Big( \Gamma_{45}-\Gamma_{67}\Big)\Gamma_8 \;\epsilon
\eea
and 
\bea
\epsilon^{abcd} F_{3a}\Gamma_{bcd} \; \epsilon = {6\over r^2} \partial_3 f y^d \Gamma_d  \; \epsilon,
\eea
where we used the projection condition (\ref{projb}) to simplify the expressions above. Employing these identities together (\ref{gammonea}) can be expressed as follows
\bea
\Gamma_{n=1} \, \epsilon &=&\Big[{2\over \sqrt{D}} (r^2+X_8^2) {\partial_r X_8\over r} {f\over r^2} y_d \sigma_{1}\otimes \Gamma_{0123} \Gamma_{d8} + {1\over \sqrt{D}}(r^2+X_8^2)  \Big( {f\over r^2}+ {\partial_r f\over2 r}\Big) \sigma_1\otimes \Gamma_{0123}(\Gamma_{45}-\Gamma_{67})\no\\
&&+  {1\over \sqrt{D}} \partial_3 X_8 \Big( {f\over r^2}+ {\partial_r f\over 2r}\Big)  \sigma_1\otimes \Gamma_{0128}(\Gamma_{45}-\Gamma_{67})+ {1\over \sqrt{D}} {1\over r^2}\partial_3 f y^d  \sigma_1\otimes \Gamma_{012d} \no\\
&&- {1\over \sqrt{D}} {1\over 2}{\partial_r X_8\over r} \partial_3 f  \sigma_1\otimes \Gamma_{0128}(\Gamma_{45}-\Gamma_{67})\Big] \epsilon.
\eea
This can be simplified by using (\ref{proja}), 
\bea
\Gamma_{n=1} \, \epsilon &=&\Big[{2\over \sqrt{D}} (r^2+X_8^2) {\partial_r X_8\over r} {f\over r^2} \sigma_{3}\otimes y_d  \Gamma_{d8} + {1\over \sqrt{D}}(r^2+X_8^2)  \Big( {f\over r^2}+ {\partial_r f\over2 r}\Big) \sigma_3\otimes  (\Gamma_{45}-\Gamma_{67})\no\\
&&- {1\over \sqrt{D}} \partial_3 X_8 \Big( {f\over r^2}+ {\partial_r f\over 2 r}\Big)  \sigma_3\otimes \Gamma_{38}(\Gamma_{45}-\Gamma_{67})-{1\over \sqrt{D}} {1\over r^2}\partial_3 f y^d  \sigma_3\otimes \Gamma_{3d} \no\\
&&+ {1\over \sqrt{D}}{1\over 2} {\partial_r X_8\over r} \partial_3 f  \sigma_3\otimes \Gamma_{38}(\Gamma_{45}-\Gamma_{67})\Big] \epsilon.
\eea
The dependence on the $y^i$ can be removed by employing (\ref{projc}) from which the following identities can be derived
\bea
y^d \Gamma_d \Gamma_{45} \epsilon &=& x^d \Gamma_d \epsilon, \no\\
\sigma_3 \otimes y^d \Gamma_{d8}\epsilon &=&-  x^d  \;1_2\otimes \Gamma_{d3}\epsilon, \no\\
\sigma_3 \otimes y^d \Gamma_{d3}\epsilon &=& +x^d  \; 1_2\otimes \Gamma_{d8}\epsilon, \no\\
\sigma_3\otimes (\Gamma_{45}-\Gamma_{67})\epsilon &=& -2 \; 1_2 \otimes \Gamma_{38}\epsilon, \no\\
\sigma_3\otimes \Gamma_{38}(\Gamma_{45}-\Gamma_{67})\epsilon &=& 2 \; 1_2 \otimes 1 \epsilon. 
\eea
Then the $\Gamma_{n=1}$ projector becomes
\bea
\Gamma_{n=1} \, \epsilon &=&\Big[ -{2\over \sqrt{D}} (r^2+X_8^2) {\partial_r X_8\over r} {f\over r^2}  x^d  \;1_2\otimes \Gamma_{d3} - {1\over \sqrt{D}}(r^2+X_8^2)  \Big( {2f\over r^2}+ {\partial_r f\over r}\Big)  1_2 \otimes \Gamma_{38} \no\\
&& -{1\over \sqrt{D}} \partial_3 X_8 \Big( {2f\over r^2}+ {\partial_r f\over  r}\Big)  \; 1_2 \otimes 1 +{1\over \sqrt{D}} {1\over r^2}\partial_3 f  ^d  \; 1_2\otimes \Gamma_{d8} \no\\
&&+ {1\over \sqrt{D}} {\partial_r X_8\over r} \partial_3 f  \; 1_2 \otimes 1 \Big] \epsilon \label{bpsone}.
\eea

\subsection{Calculation of $\Gamma_{n=2}$}
The part of the projector which is quadratic in the field strength is given by
\bea
\Gamma_{n=2}&=& {1\over \sqrt{D}}\Big( {1\over 2}\gamma^{3abc}F_{3a}F_{bc}+{1 \over 8}\gamma^{abcd}F_{ab}F_{cd}\Big)i\sigma_2 \otimes \gamma_{01234567} \no\\
&=&  {1\over \sqrt{D}} \Big( {1\over 2} \epsilon^{abcd}   F_{3a} F_{bc} i\sigma_2 \otimes  \gamma_{012d}+ {1\over 8} \epsilon^{abcd} F_{ab}F_{cd}i\sigma_2 \otimes  \gamma_{0123}\Big) .
\eea
Expressing the world volume gamma matrices in terms of flat space ones gives
\bea
\gamma_{0123} &=& (r^2+X_8^2)^2 \Gamma_{0123} + (r^2+X_8^2) \partial_3  X_8 \Gamma_{0128} ,\no\\
\gamma_{012d}  &=& (r^2+X_8^2)  \Big(\Gamma_{012d} + {\partial_r X_8\over r} x_d  \Gamma_{0128} \Big)
\eea
and using the following relations which follow from the explicit ansatz for the gauge field (\ref{gaugef})
\bea
{1\over 8} \epsilon^{abcd} F_{ab}F_{cd} &=& -{2\over r^3} f \partial_r f, \no\\
{1\over 2}  \epsilon^{abcd} F_{3a }F_{bc} \Gamma_d &=&+{2\over r^4} f \partial_3 f\;  x^d  \Gamma_d, \no\\
{1\over 2} \epsilon^{abcd} F_{3a} F_{bc} x_d &=& +{2\over r^2} f \partial_3 f,
\eea
we obtain
\bea
\Gamma_{n=2} \, \epsilon &=& {1\over \sqrt{D}}\left\{  -{2\over r^3} (r^2+X_8^2)^2   f \partial_r f   \;i\sigma_2 \otimes  \Gamma_{0123}  -  {2\over r^3} (r^2+X_8^2)   f \partial_r f   \partial_3 X_8   \;i\sigma_2 \otimes \Gamma_{0128} \right. \no\\
&& \quad \left.  +{2\over r^4} (r^2+X_8^2) f \partial_3 f\;  x^d \; i\sigma_2\otimes   \Gamma_{012d}  + {2\over r^3} (r^2+X_8^2) f \partial_3 f\; \partial_r X_8   \;i\sigma_2 \otimes  \Gamma_{0128} \right\} \epsilon .\nonumber\\
\eea
Using the projector (\ref{proja}) the expression can be simplified further to give
\bea
\Gamma_{n=2} \, \epsilon &=& {1\over \sqrt{D}}\left\{  -{2\over r^3} (r^2+X_8^2)^2   f \partial_r f   \;  {1_2\otimes 1} + {2\over r^3} (r^2+X_8^2)   f \partial_r f   \partial_3 X_8  \; 1_2\otimes \Gamma_{38} \right. \no\\
&& \quad \left.  +{2\over r^4} (r^2+X_8^2) f \partial_3 f\;  x^d \;  1_2\otimes \Gamma_{d3}  - {2\over r^3} (r^2+X_8^2) f \partial_3 f\; \partial_r X_8   \; 1_2\otimes \Gamma_{38} \right\} \epsilon.\label{bpstwo}\nonumber\\
\eea
\subsection{Final form of the $\Gamma$ projector}
The  expressions for $\Gamma_{n=0,1,2}$ in are evaluated in append \ref{appendb}. 
The final result for the projector $\Gamma= \Gamma_{n=0}+\Gamma_{n=1}+\Gamma_{n=2}$ from (\ref{bpszero}), (\ref{bpsone}) and (\ref{bpstwo}) one gets the final form of the $\Gamma$ projector.
\bea\label{gammafullappend}
\Gamma \, \epsilon &=& {1\over \sqrt{D}} \Big\{ 1+ {\partial_r X_8 \over r}\partial_3 f -{2\over r^3} (r^2+X_8^2)^2 f \partial_r f - \partial_3 X_8 \big({2f\over r^2}+ {\partial _r f \over r}\big) \Big\} 1_2\otimes 1 \, \epsilon \no \\
&+& {(r^2+X_8^2)\over \sqrt{D}} \Big\{  {-\partial_3 X_8\over (r^2+X_8^2)^2} -  \big({2f\over r^2}+ {\partial _r f \over r}\big) +{2 f\over r^3}  \big( \partial_r f \partial_3 X_8 - \partial_3 f \partial_r X_8\big)\Big\} 1_2\otimes \Gamma_{38} \, \epsilon  \no\\
&+&  {2(r^2+X_8^2) f \over r^3  \sqrt{D}} \Big\{ {\partial_3 f\over r}-\partial_r X_8\Big\} 1_2\otimes x^d \Gamma_{d3} \, \epsilon   +{1\over \sqrt{D}}{1\over r}\Big\{ {\partial_3 f\over r}-\partial_r X_8\Big\} 1_2\otimes x^d \Gamma_{d8} \, \epsilon. \no\\
\eea

\end{document}